\documentclass[manuscript]{emulateapj}
\usepackage{subfigure}

\shorttitle{Chromospheric cannonballs on the Sun}
\shortauthors{Yang et al.}

\journalinfo{Accepted for publication in ApJL}
\submitted{ApJL, accepted 2019 June 25}

\begin{document}

\title{Chromospheric cannonballs on the Sun}

\author{Shuhong Yang\altaffilmark{1,2}, Jun Zhang\altaffilmark{1,2}, Xiaohong Li\altaffilmark{1,2}, Zhong Liu\altaffilmark{3}, Yongyuan Xiang\altaffilmark{3}}

\altaffiltext{1}{CAS Key Laboratory of Solar Activity, National Astronomical Observatories, Chinese Academy of Sciences, Beijing 100101, China; (shuhongyang@nao.cas.cn; zjun@nao.cas.cn)}

\altaffiltext{2}{School of Astronomy and Space Science, University of Chinese Academy of Sciences, Beijing 100049, China}

\altaffiltext{3}{Fuxian Solar Observatory, Yunnan Observatories, Chinese Academy of Sciences, Kunming 650011, China}

\begin{abstract}

In the highly dynamic chromosphere, there exist many kinds of small-scale activities, such as spicules, surges, and Ellerman bombs. Here, we report the discovery of a new phenomenon in the chromosphere observed with the New Vacuum Solar Telescope at the \emph{Fuxian Solar Observatory}. In the high tempo-spatial resolution H$\alpha$ images, some dark or bright structures are found to fly along the curved trajectory, looking like cannonballs. Their average size, mass, and velocity are about 1.5 $\times$ 10$^{9}$ km$^{3}$, 1.5 $\times$ 10$^{8}$ kg, and 56 km s$^{-1}$, respectively.  In the simultaneous (extreme-)ultraviolet images obtained by the \emph{Solar Dynamics Observatory}, these cannonballs appear as brighter features compared to the surrounding area, implying that there exists some kind of heating during this process. The photospheric magnetograms show the magnetic flux emergence and interaction with the pre-existing fields. These observations reveal that the cannonballs are chromospheric material blobs launched due to the magnetic reconnection between emerging magnetic flux and the pre-existing loops.

\end{abstract}

\keywords{magnetic reconnection --- Sun: activity --- Sun: chromosphere --- Sun: magnetic fields}

\section{Introduction}

The solar chromosphere is the necessary layer of energy transport from the photosphere to the corona. Thus the chromosphere of the Sun plays an important role in coronal heating and solar wind acceleration. As the temperature minimum layer of the solar atmosphere, the chromosphere has been extensively imaged in H$\alpha$ wavelength with ground-based and space-based instruments for many years. Dark fibrils are ubiquitous in the solar chromosphere, and they are filled with cool and dense material (Aschwanden et al. 2016). Comparing the fibrils in the H$\alpha$ images with the extrapolated potential fields from the photospheric magnetograms, the chromospheric magnetic fields are found to be significantly nonpotential (Woodard \& Chae 1999). The free energy stored in nonpotential fields can be released to power plentiful solar activities in the chromosphere (Moore et al. 2010; Sterling et al. 2015; Yang et al. 2015; Ni et al. 2016; Xue et al. 2016; Yang \& Xiang 2016). Spicules, appearing as thin and dynamic extrusions if observed at the solar limb, are ubiquitous in the chromosphere (Sterling 2000; De Pontieu et al. 2004). De Pontieu et al. (2007) found that some spicules are driven by shock waves along magnetic field lines, and some ones sending material through the chromosphere at higher speeds are caused by magnetic reconnection. Surges, jet-like structures observed in H$\alpha$ wavelength, are always associated with magnetic flux emergence or cancellation in observations, and are theoretically deemed to result from magnetic reconnection (Yokoyama \& Shibata 1995; Chae et al. 1999; Shibata et al. 2007). Ellerman bombs are small-scale short-lived brightenings in both wings of the H$\alpha$ line (Ellerman 1917; Fang et al. 2006; Tian et al. 2016). Their triggering mechanism is suggested to be a reconnection process caused by the low-lying magnetic fields in the chromosphere (Georgoulis et al. 2002; Reid et al. 2016). The MHD simulations also show that, when the emerging magnetic flux expands in the atmosphere, magnetic reconnection takes place and consequently heats the local dense plasma (Isobe et al. 2007; Archontis \& Hood 2009).

With the improvement of observational instrument, it is a good opportunity for us to further study the fine-scale chromospheric structures. In the present Letter, we report the discovery of a new phenomenon in the chromosphere, i.e., flying cannonball-like structures (hereafter cannonballs), based on the high tempo-spatial resolution H$\alpha$ observations from the New Vacuum Solar Telescope (NVST; Liu et al. 2014) of China.

\begin{figure*}
\centering
\includegraphics
[bb=56 190 505
643,clip,angle=0,width=0.8\textwidth]{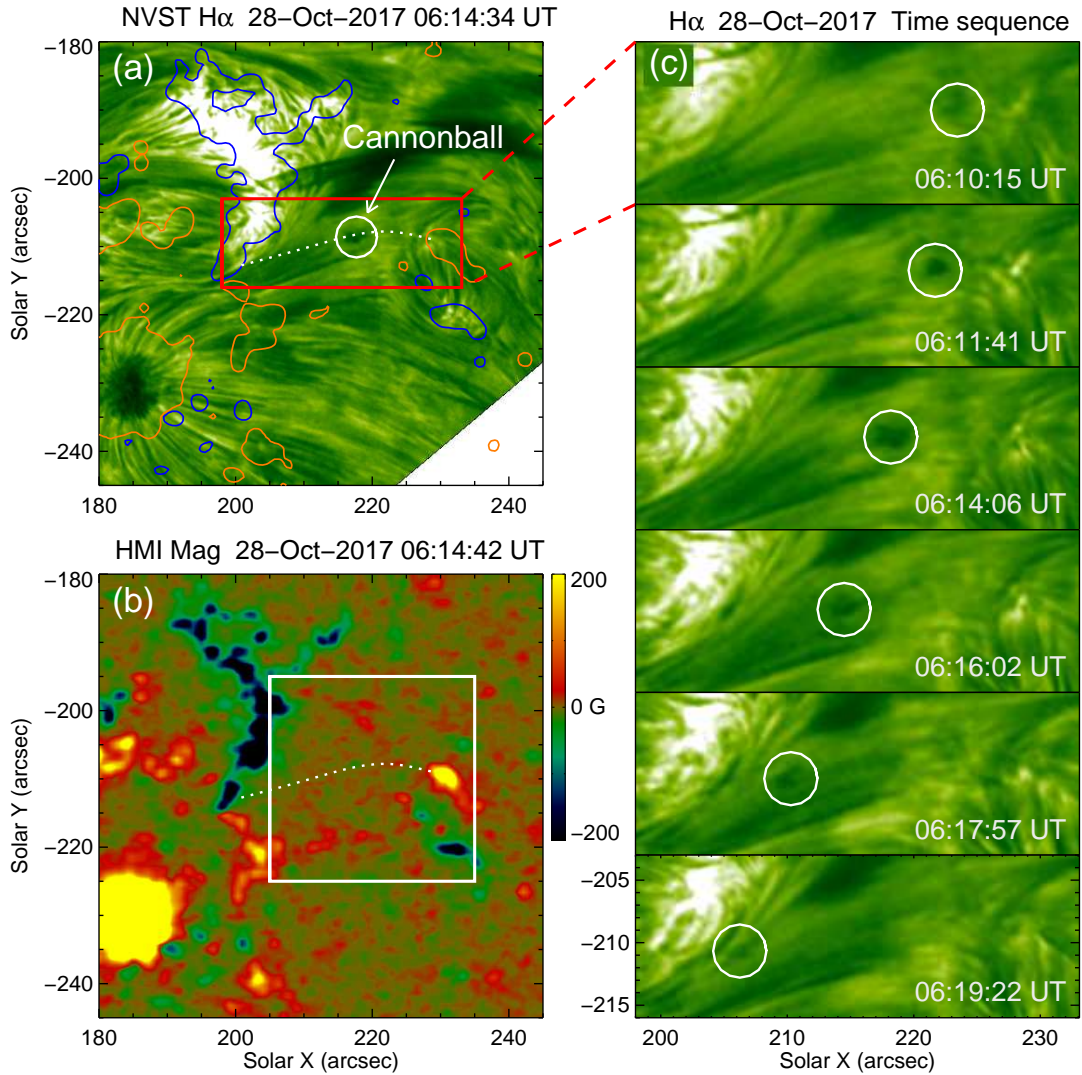} \caption{ \emph{Panels (a)-(b)}: NVST H$\alpha$ image and HMI LOS magnetogram displaying the chromospheric cannonball-like structure and the underlying photospheric magnetic environment, respectively. \emph{Panel (c)}: Time sequence of H$\alpha$ images showing the movement of the cannonball along a curved trajectory.
The brown and blue solid curves in panel (a) are the contours of the photospheric magnetic fields at + 30 G and - 30 G, respectively. The white circles and dotted curves outline the cannonball and its flying trajectory, respectively. The square in panel (b) denotes the FOV of Figures 2(a)-(c). \protect\\\emph{One animation (Movie1.mp4) of this figure is available.} \label{fig}}
\end{figure*}

\begin{figure*}
\centering
{\subfigure{\includegraphics[bb=107 297 489 535,clip,angle=0,width=0.84\textwidth]{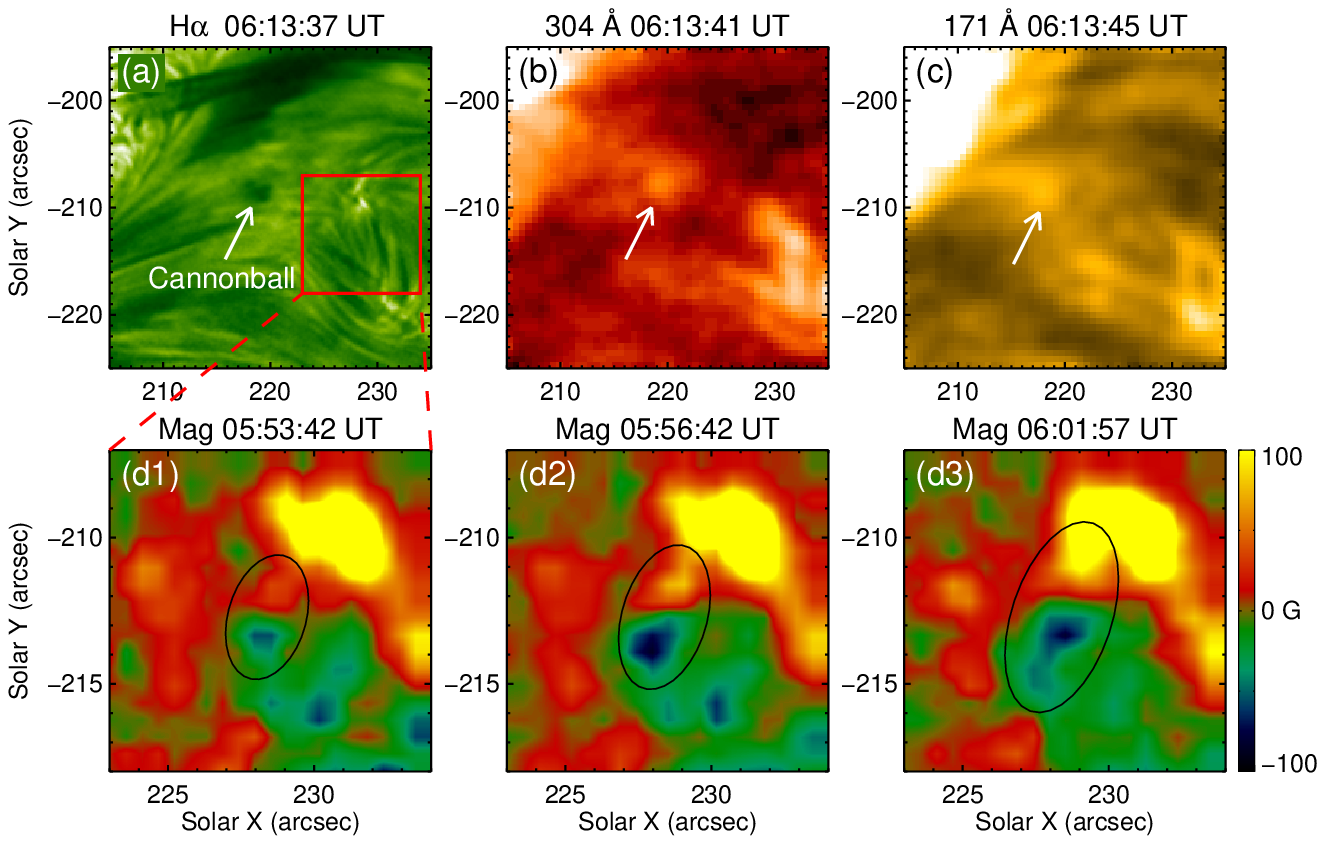}} \quad
\subfigure{\includegraphics[bb=149 344 455 576,clip,width=0.65\textwidth]{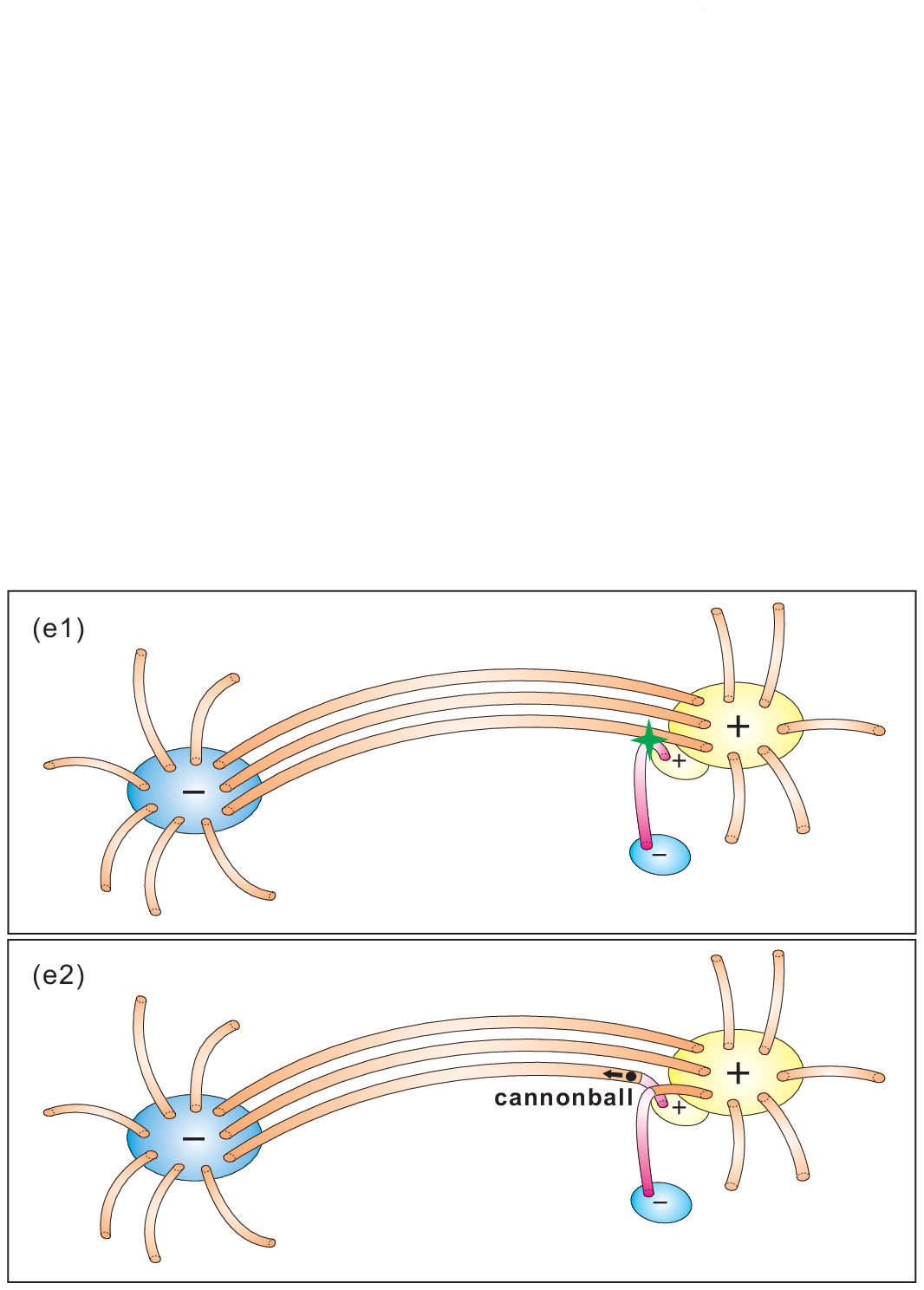}}}
\caption{\emph{Panels (a)-(c)}: NVST H$\alpha$, AIA 304 {\AA} and 171 {\AA} images displaying the different appearance of the cannonball (denoted by the arrows). \emph{Panels (d1)-(d3)}: HMI LOS magnetograms showing the emergence of a bipole (outlined by the ellipses) at the origination site of the cannonball. \emph{Panels (e1)-(e2)}: Schematic drawings illustrating the magnetic reconnection resulting in the formation of cannonballs. \protect\\\emph{Two animations (Movie2.mp4 \& Movie3.mp4) of this figure are available.} \label{fig}}
\end{figure*}

\begin{figure*}
\centering
{\subfigure{\includegraphics[bb=44 298 546 541,clip,angle=0,width=\textwidth]{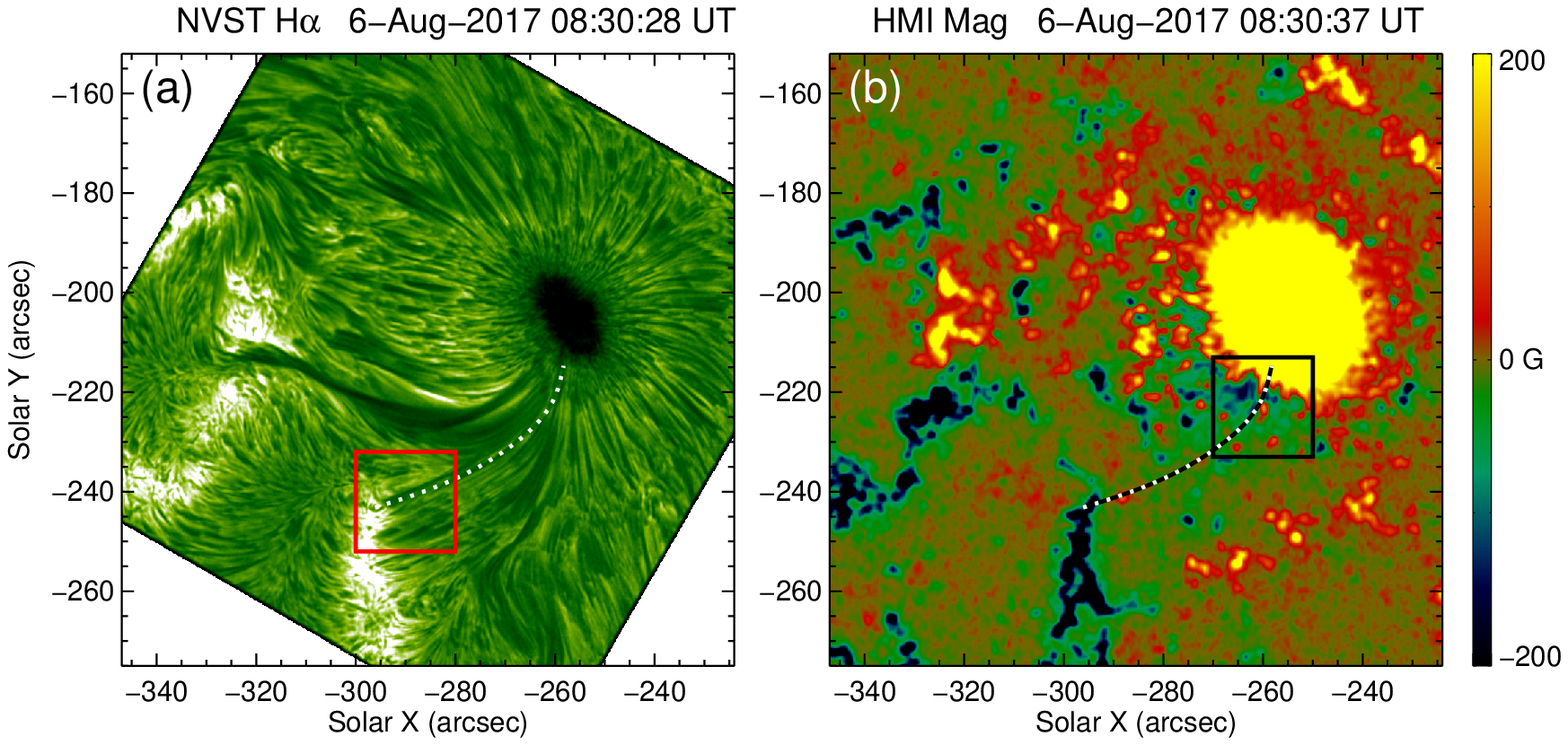}} \quad
\subfigure{\includegraphics[bb=44 345 551 483,clip,angle=0,width=\textwidth]{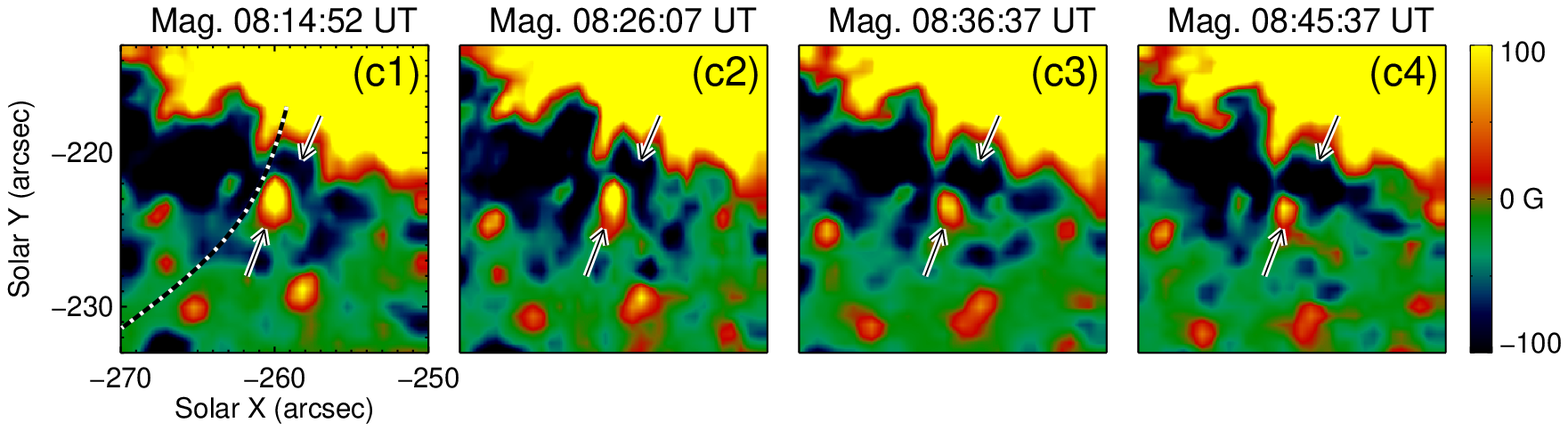}}}
\caption{\emph{Panels (a)-(b)}: Overview of the H$\alpha$ observation and the photospheric magnetic field, respectively. The dotted curves delineate the fibril along which the cannonball flew. The box in panel (b) outlines the FOV of panels (c1)-(c4), and the box in panel (a) outlines the FOV of Figure 4. \emph{Panels (c1)-(c4)}: Sequence of HMI magnetograms showing the magnetic field evolution at the initiation site of the cannonball. The upper arrows indicate the emerging negative field and the lower arrows denote the disappearing positive field. \protect\\\emph{One animation (Movie4.mp4) of this figure is available.} \label{fig}}
\end{figure*}

\begin{figure*}
\centering
{\subfigure{\includegraphics[bb=44 302 525 541,clip,angle=0,width=\textwidth]{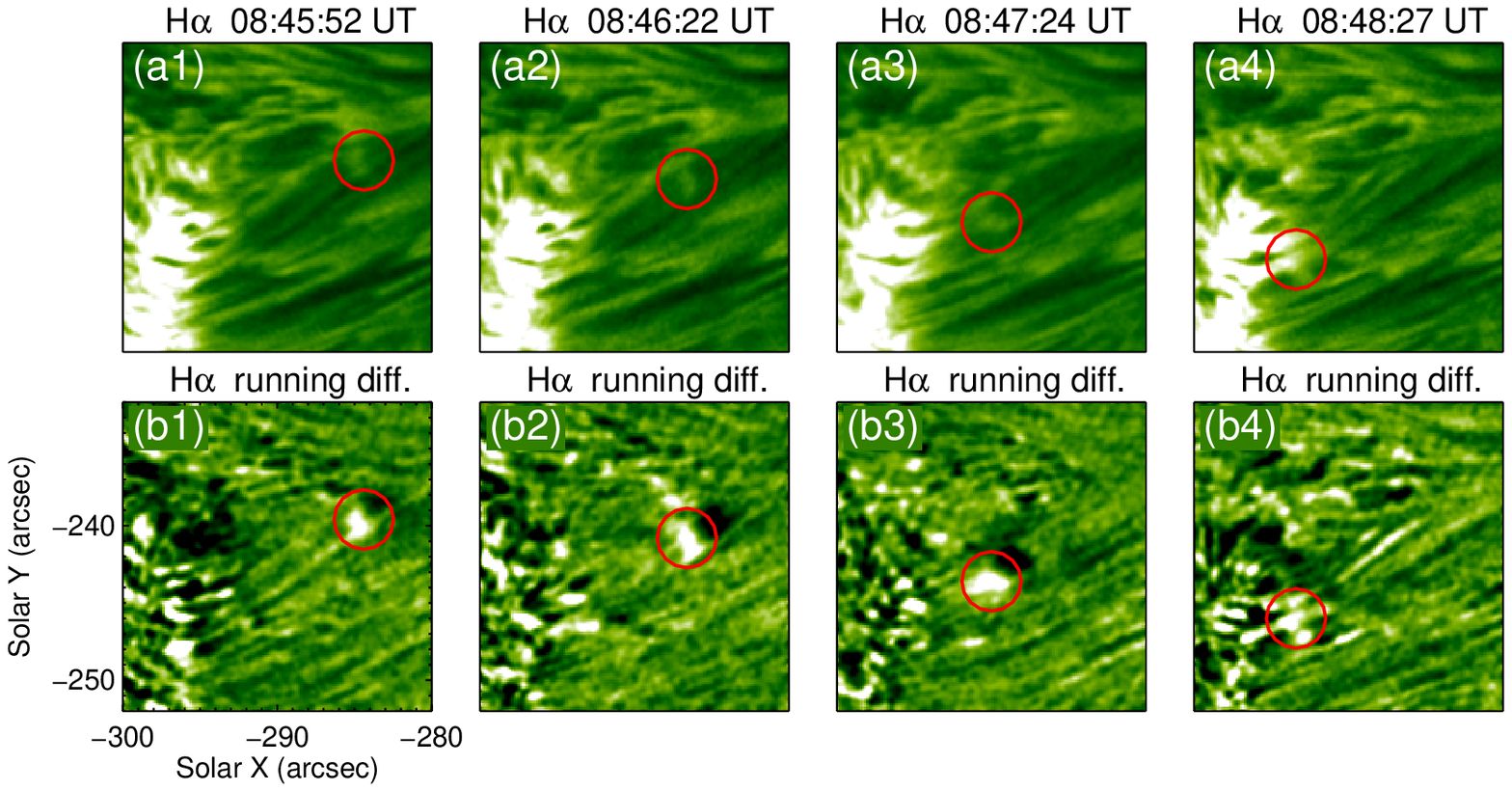}} \quad
\subfigure{\includegraphics[bb=76 345 513 496,clip,angle=0,width=0.9\textwidth]{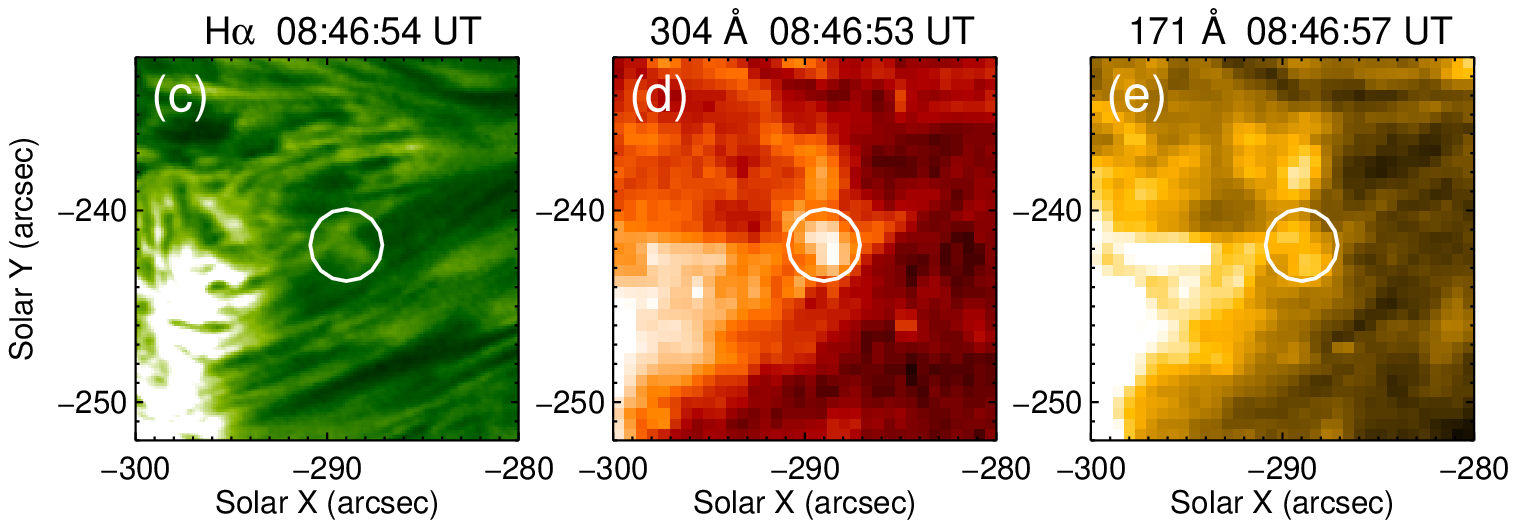}}}
\caption{\emph{Panels (a1)-(a4)}: H$\alpha$ Level 1+ images displaying the flying cannonball (outlined by the circles). \emph{Panels (b1)-(b4)}: H$\alpha$ running difference images corresponding to the upper panels. \emph{Panels (c)-(e)}: Multi-wavelength appearance of the cannonball in H$\alpha$, 304 {\AA}, and 171 {\AA}, respectively. \protect\\\emph{Two animations (Movie5.mp4 \& Movie6.mp4) of this figure are available.} \label{fig}}
\end{figure*}

\begin{figure*}
\centering
\includegraphics
[bb=51 251 524 581,clip,angle=0,width=0.95\textwidth]{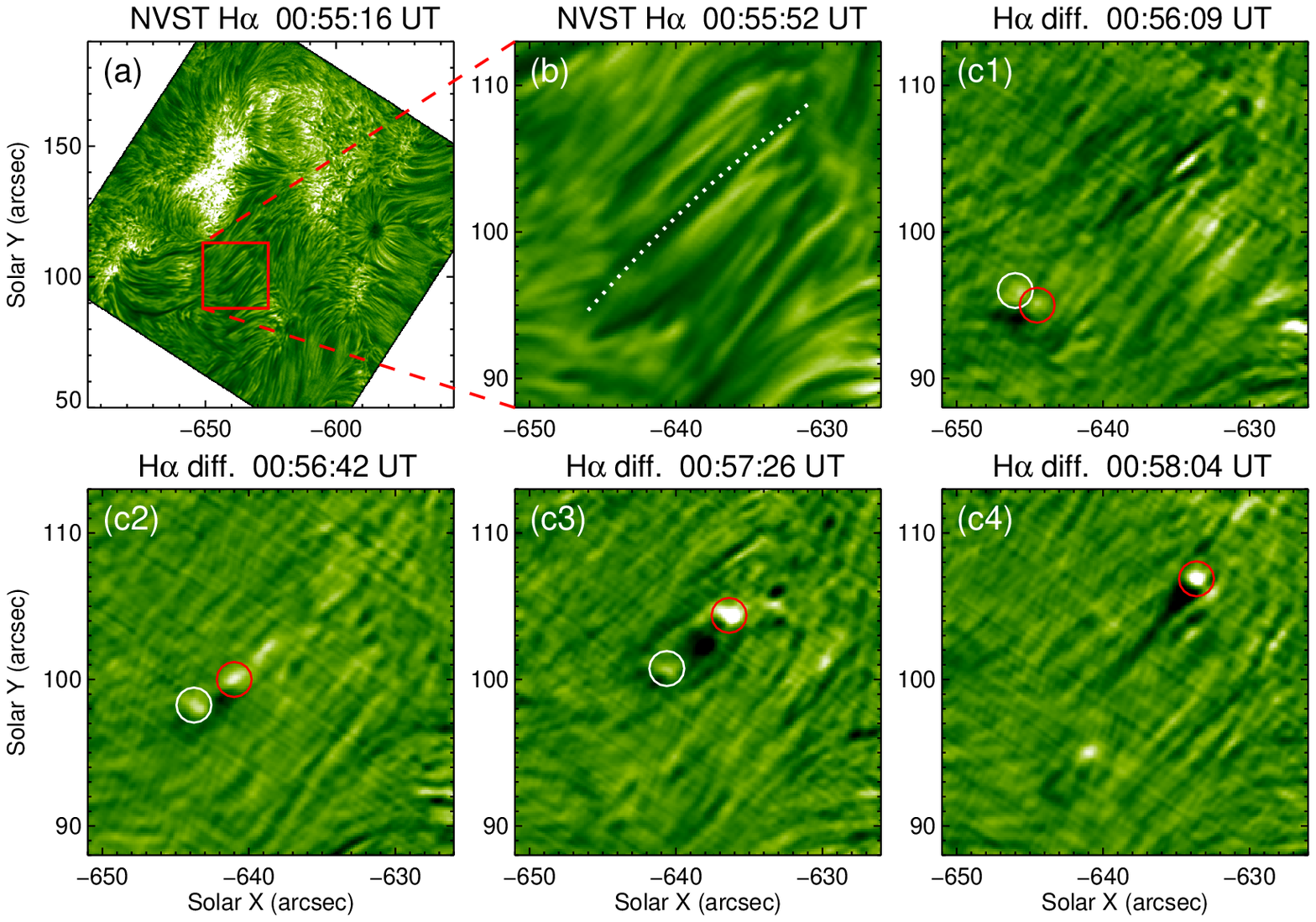} \caption{ \emph{Panel (a)}: Overview of the H$\alpha$ observation on 2018 May 11. \emph{Panel (b)}: H$\alpha$ Level 1+ image of a sub-FOV where the dotted curve delineates
the trajectory of two cannonballs. \emph{Panels (c1)-(c4)}: H$\alpha$ running difference images showing the movement of two cannonballs outlined with the white and red circles, respectively. \protect\\\emph{One animation (Movie7.mp4) of this figure is available.}
\label{fig}}
\end{figure*}

\section{Observations and data analysis}

The NVST with a clear aperture of 985 mm is the primary facility of the \emph{Fuxian Solar Observatory} operated by Yunnan Observatories. H$\alpha$ channel is used to image the highly dynamic solar chromosphere at high tempo-spatial resolution. The H$\alpha$ filter is a tunable Lyot filter with a bandwidth of 0.25 {\AA}, which can scan spectra in the $\pm$ 5 {\AA} range with a step size of 0.1 {\AA}. The present study is mainly based on three events with the H$\alpha$ line-center (6562.8 {\AA}) observations, i.e., 2017 October 28 05:43-08:01 UT, 2017 August 06 08:00-10:06 UT, and 2018 May 11 00:50-05:02 UT. Their time cadences are 29 s, 31 s, and 9 s, respectively. The fields of view (FOVs) are 126{\arcsec} $\times$ 126{\arcsec} with the pixel size of 0.{\arcsec}136. The Level 0 data are first calibrated to Level 1, including dark current subtraction and flat field correction. Then the calibrated Level 1 images are reconstructed to Level 1+ by speckle masking (Weigelt 1977; Lohmann et al. 1983).

The simultaneous line-of-sight (LOS) magnetograms from the Helioseismic and Magnetic Imager (HMI; Scherrer et al. 2012) and the multi-wavelength images from the Atmospheric Imaging Assembly (AIA; Lemen et al. 2012) on board the \emph{Solar Dynamics Observatory} (\emph{SDO}; Pesnell et al. 2012) are also used. The HMI LOS magnetograms have a cadence of 45 s and a spatial sampling of 0.{\arcsec}5 pixel$^{-1}$, and the AIA (extreme-)ultraviolet (EUV/UV) full-disk images have a cadence of 12/24 s and a pixel size of 0.{\arcsec}6. Here, we mainly present the AIA observations in 304 {\AA} (UV) and 171 {\AA} (EUV) lines, in which the cannonballs have the conspicuous counterparts. All the HMI and AIA data are calibrated to Level 1.5 with the standard routine under the Solar Software package. Then the data on October 28 and August 06 are de-rotated differentially to the reference times of 06:00 UT and 08:46 UT, respectively. The NVST H$\alpha$ images are coaligned with the \emph{SDO} data using the cross-correlation method according to specific features.

\section{Results}

The FOV of the NVST H$\alpha$ observations on 2017 October 28 is centered at (190", -200"). Figure 1(a) displays the appearance of a sub-FOV H$\alpha$ image at 06:14:34 UT where a cannonball is enclosed by the white circle. The dotted curve delineates the trajectory, along which the cannonball moved from the west to the east. The origination and landing sites of the cannonball correspond to the positive and negative magnetic patches, respectively (panel (b)). To see the movement well, we show a sequence of enlarged H$\alpha$ images in Figure 1(c), and the cannonball at each time is outlined by a white circle. At 06:10:15 UT, the cannonball can be well identified as a dark structure. One and a half minutes later, the cannonball was more conspicuous and moved to a new site to the east (see also Movie 1). In the following seven and a half minutes, the cannonball went on flying with the average velocity of about 23.4 km s$^{-1}$.

Since the cannonball appears as a quasi-circular feature in the H$\alpha$ images, it can be assumed to be a ball in three dimensions. If the diameter of a ball measured in the observational image is \emph{D}, then the size (i.e., volume) \emph{V} is $4/3 \times \pi \times (D/2)^{3}$. In the 2017 October 28 event, the \emph{D} of the cannonball is 1270 km, so the size of the cannonball is $1.07 \times 10^{9}$ km$^{3}$. The density of the cannonball is considered to be 1 $\times$ 10$^{-10}$ kg m$^{-3}$, i.e., the density of the chromosphere (Beckers 1968). Thus the cannonball has the mass of 1.07 $\times$ 10$^{8}$ kg.

In order to study the temperature character of the cannonball, we examine the multi-wavelength images (see Movie 2). We find that, for the dark cannonball in the H$\alpha$ image (denoted by the arrow in Figure 2(a)), there are clear bright counterparts in 304 {\AA} (panel (b)) and 171 {\AA} (panel (c)) images. In the UV (304 {\AA}) and EUV (171 {\AA}) lines, the emission of these counterparts is significantly higher than the surrounding area. Since magnetic field evolution always plays a crucial role in solar active events, we also examine the photospheric magnetograms at the origination site of the cannonball. A time sequence of HMI magnetograms during a 15 min period before the cannonball onset are shown in the lower panels of Figure 2 (see also Movie 3). The positive polarity appears as the red-yellow patches, and the negative polarity as the green-blue ones. At 05:53 UT, a faint bipole (outlined by the ellipse in panel (d1)) began to emerge at the south-east location of the pre-existing positive magnetic field. Three minutes later, the bipole grew much larger (see panel (d2)). Meanwhile, the bipolar patches with opposite polarities separated from each other, and the positive patch moved to and merged with the pre-existing positive field (panel (d3)).

Different from the above event, the NVST H$\alpha$ images on 2017 August 6 present a bright cannonball instead of a dark one. Figure 3(a) shows the overview of the H$\alpha$ image where there are many long fibrils extending from the sunspot to the remote region. The dotted curve delineates a fibril along which the cannonball moved. In the HMI photospheric magnetogram (panel (b)), the trajectory of the cannonball is also overlaid. The cannonball originated from the penumbral positive field and flew to the south-east enhanced negative field. The magnetic field evolution at the origination site of the cannonball is outlined by the black square in panel (b) and displayed in panels (c1)-(c4). We can see that the negative polarity patch pointed by the upper arrows continued growing, while the positive polarity patch denoted by the lower arrows shrank significantly (see also Movie 4). These changes indicate the existence of magnetic emergence and cancellation at the origination site of the cannonball.

In the H$\alpha$ images in Figure 4, the cannonball can be directly identified (top panels; see also Movie 5). The cannonball outlined by the red circles moved from higher-right to the lower-left. The size of the cannonball is about 1.74 $\times$ 10$^{9}$ km$^{3}$, and the mass is estimated to be 1.74 $\times$ 10$^{8}$ kg. Since the cannonball is somewhat faint in the Level 1+ data, we use the H$\alpha$ running difference images to show it (panels (b1)-(b4)). We can see that the cannonball is more conspicuous in the running difference images, where a bright feature followed by an adjacent dark one represents the cannonball. Its average movement velocity is about 50.0 km s$^{-1}$. In addition, we check the multi-wavelength images to study the counterpart of the cannonball. The bright chromospheric cannonball in H$\alpha$ image (panel (c)) also appears as a bright feature in both AIA 304 {\AA} (panel (d)) and 171 {\AA} (panel (e)) images (see also Movie 6). Particularly, in the 304 {\AA} images, the cannonball's brightness relative to the surrounding area is more prominent, compared with its appearance in the 171 {\AA} images.

Figure 5(a) shows the overview of the H$\alpha$ image on 2018 May 11, when two bright cannonballs were observed (see also Movie 7). The cannonballs flew to the north-west along the dotted curve in panel (b). Because the signals of the cannonballs are weak in the H$\alpha$ Level 1+ images, we study them in the running difference images (see panels (c1)-(c4)), where a bright feature followed by an adjacent dark one represents a cannonball. These two cannonballs (outlined by the white and red circles respectively) originated side by side from the south-east end of the dark fibrils (see panel (c1)). The left cannonball flew to the north-west with a mean velocity of 60 km s$^{-1}$, and the right one was much faster with the average velocity of 90 km s$^{-1}$. Their average size and mass are 1.65 $\times$ 10$^{9}$ km$^{3}$ and 1.65 $\times$ 10$^{8}$ kg, respectively. When we check the underlying magnetograms, we find that the cannonballs were launched from the intranetwork magnetic field with mixed polarities and flew to the network field with negative field. The emergence and cancellation of small-scale intranetwork fields are widespread, which can contribute to the formation and launch of cannonballs.

\section{Conclusions and discussion}

Based on the H$\alpha$ observations from the NVST, we find a new phenomenon in the highly dynamic chromosphere. Some dark or bright structures flew along the curved trajectory, looking like cannonballs. The average size, mass, and velocity of the cannonballs are about 1.53 $\times$ 10$^{9}$ km$^{3}$, 1.53 $\times$ 10$^{8}$ kg, and 55.9 km s$^{-1}$, respectively. In the simultaneous UV and EUV images from the AIA, these cannonballs appear as brighter structures compared to the surrounding environment. The photospheric magnetograms from the HMI show the emergence of magnetic flux and the consequent interaction with the pre-existing fields.

The reasons why we call these observed structures cannonballs are as below. (1) They have similar morphology to cannonballs since they appear as quasi-spherical blobs in observations. (2) Their movement consists of a rising stage and then a falling stage in height, as they originate from one end of the curved field lines and move toward the other end, which is also the general behavior of flying cannonballs.

To illustrate the cannonball formation, we propose a cartoon as shown in Figures 2(e1)-(e2). Initially, there exists a positive patch connected with remote negative field by large-scale loops (panel (e1)). A bipole emerges at the adjacency to the pre-existing positive patch, and its two opposite polarities are connected by small-scale loops. As the emerging bipolar patches separate, the small-scale loops rise towards the overlying loops. Eventually, the newly emerged small-scale loops meet and reconnect with the overlying large-scale field lines. Due to the reconnection, the chromospheric material is expelled from the reconnection site and flies along the large-scale loops, forming a cannonball (panel (e2)).

In the observations, both the dark and bright cannonballs have similar behaviors. For these H$\alpha$ cannonballs, there are prominent bright counterparts in the UV/EUV lines, implying the existence of some kind of heating mechanism. We suggest that magnetic reconnection may result in the heating of the cannonballs as observed. Previous studies have shown that magnetic flux emergence and cancellation are often associated with magnetic reconnection (Zhang et al. 2001; Yang et al. 2011; Hansteen et al. 2017). As shown in Figures 2 and 3, we indeed observed the magnetic emergence/cancellation in the photosphere at the origination sites. For the 2018 May 11 event, the initiation site of the cannonballs is located in the intranetwork region, where magnetic reconnection between emerging flux with mixed polarities is ubiquitous. When magnetic reconnection takes place, magnetic energy is converted into thermal energy and kinetic energy (Zweibel \& Yamada 2009; Cirtain et al. 2013; Wyper et al. 2017). Therefore the cannonballs can be consequently launched and heated. However, for the heating of the cannonballs, another possibility could not be excluded, i.e., being heated by the shock front of the flying cannonballs. The flying cannonballs can play an importance role in transferring material and energy in the solar atmosphere.

We should note that reconnection physics under solar conditions has not been understood well enough to state with any certainty the thermal physics. Magnetic reconnection is a rapid, dynamic change in topology leading to bulk flows, but the temperature of the gas is highly uncertain. In the solar chromosphere it may be simpler than in the corona, considering that the reconnecting Lorentz forces act only on ions and that the chromosphere is almost fully neutral which will lead to ion-neutral heating. However, this is just one aspect of magnetic reconnection under chromospheric conditions.

The highly dynamic structures (such as fast-moving bright blobs) in the chromosphere were also revealed in the Swedish 1 m Solar Telescope (SST) observations (van Noort \& Rouppe van der Voort 2006). However, the cannonballs are different from those previous observed structures. The SST observed the bright blobs, while the NVST observed both the dark and bright features. The SST bright blobs are high-velocity (up to 240 km s$^{-1}$) structures, while the NVST cannonballs are low-velocity features with the average velocity of about 37 km s$^{-1}$. For the SST blobs, there appear many conspicuous brightenings at the origination sites, which cannot be identified during the initiation of the NVST cannonballs. Moreover, there are significant topological changes in the SST observations, while the chromospheric structures observed with the NVST are generally quiet and stable. Therefore, the SST blobs are associated with eruptive events, and the NVST cannonballs only result from small-scale magnetic activity in the chromosphere.

As described in the previous studies (e.g., Schrijver 2001; Antolin et al. 2010; Antolin \& Rouppe van der Voort 2012), when the coronal matter cools down catastrophically due to the thermal instability, the condensations fall from coronal heights along coronal loops to the footpoints under the action of the solar gravity, forming ``coronal rain''. While in our observations, especially clearly displayed in Figures 1 and 5, cannonballs are chromospheric material which originated at one end of the magnetic loops and flew to the other end. Therefore, the cannonballs we observed are different from frequently studied coronal rain.

Both the cannonballs and type II spicules are thought to be the consequences of magnetic reconnection, which can be used to interpret the fact that they have the similar speeds. However, they are different phenomena, not merely in the aspect that the cannonballs occur along locally closed field lines instead of open ones. As revealed by the previous studies (e.g., De Pontieu et al. 2007; Pereira et al. 2012), type II spicules are thin jet-like features (no more than 200 km in width and at least several Mm in length). The cannonballs shown in the present paper are quasi-spherical, and much wider and shorter than type II spicules. From these three events studied here, we expect to find about 40 cannonballs on the Sun at any given time. This number is much lower than that of type II spicules which is surmised to be several tens per supergranule (e.g., Moore et al. 2011; Sekse et al. 2012).

Then an interesting question is raised: why are the cannonballs like ``blobs" of fluid instead of ``jets" that are commonly thought to be type II spicules? The cannonballs appear much more morphologically like the coronal rain than jets. How can this be? We hope the answer can be given in the near future through further studies.

\acknowledgments {We are grateful to the referee for the constructive comments and valuable suggestions. The data are used courtesy of NVST, HMI, and AIA science teams. This work is supported by the National Natural Science Foundations of China (11673035, 11790304, 11533008, 11790300), Key Programs of the Chinese Academy of Sciences (QYZDJ-SSW-SLH050), and the Youth Innovation Promotion Association of CAS (2014043). \\ }

{}

\clearpage

\end{document}